\documentclass[letterpaper, 10 pt, conference]{ieeeconf}  

\IEEEoverridecommandlockouts                              
\usepackage{url}
\usepackage{graphics} 
\usepackage{epsfig} 
\usepackage{mathptmx} 
\usepackage{times} 
\usepackage{amsmath} 
\usepackage{amssymb}  
\usepackage{tikz}
\usepackage[caption=false,font=footnotesize]{subfig}
\tikzset{badge/.style={fill=black!75, text=white, rounded corners=2pt,
    inner sep=1.5pt, minimum height=0.34cm, minimum width=0.34cm,
    font=\footnotesize\bfseries}}
\newcommand{\blk}[1]{\,\tikz[baseline=(b.base)]\node[badge](b){#1};}
\usetikzlibrary{shadows.blur}
\usetikzlibrary{shapes.geometric}
\usetikzlibrary{arrows.meta, positioning}
\usetikzlibrary{matrix}
\usepackage{booktabs}
\usepackage{cite}

\title{\LARGE \bf
An Integrated Architecture for Data-Efficient Real-Time Neural Imitation of Nonlinear Time-Optimal Model Predictive Control
}

\author{Maarten De Baecke$^{1}$, Flavia Sofia Acerbo$^{1}$, Andrea Giusti$^{2}$, and Wilm Decré$^{1}$%
\thanks{This work was supported by Flanders Make through the SBO project LearnOpTra (Learning meets optimisation for robust and multimodal trajectory planning and control).}%
\thanks{$^{1}$Maarten De Baecke, Flavia Sofia Acerbo, and Wilm Decré are with the Department of Mechanical Engineering, MECO Research Team and Flanders Make@KU Leuven, Leuven, Belgium (e-mail: \{maarten.debaecke, flaviasofia.acerbo, wilm.decre\}@kuleuven.be).}%
\thanks{$^{2}$Andrea Giusti is with Flanders Make vzw, Lommel, Belgium (e-mail: andrea.giusti@flandersmake.be\}}%
}

\begin{document}

\bstctlcite{IEEEexample:BSTcontrol}

\maketitle
\thispagestyle{empty}
\pagestyle{empty}

\begin{abstract}
Nonlinear time-optimal model predictive control (TOMPC) enables highly performant control of constrained systems, but its computational cost prohibits real-time use on fast dynamical systems. Neural network imitation of the TOMPC control law can reduce online computational costs, yet existing work mostly targets quadratic cost formulations in simulation and addresses the challenges of neural controllers in isolation. This work imitates a long-horizon nonconvex TOMPC with hard terminal constraints, whose high computational cost limits training-data generation. We present (i) \emph{self-informed sampling (SIFS)}, a data-efficient expert-sampling strategy that uses a learned time-to-go to query the expert only where the policy deviates from time-optimal behavior, (ii) a complete training and control architecture that jointly addresses dataset correlation, suboptimal expert labels, covariate shift, constraint satisfaction and model mismatch, and (iii) real-world validation on a Furuta pendulum. SIFS matches standard dataset aggregation performance in roughly one-third of the training time and with less than half the training data. The resulting controller matches the region of attraction of the expert TOMPC while raising the control frequency from 1.4~Hz to 1~kHz. On hardware, the controller swings up and stabilizes the pendulum along trajectories closely matching the expert TOMPC solution.
\end{abstract}

\section{INTRODUCTION} 
Model predictive control (MPC) has emerged as a powerful and widely applicable control strategy. In particular, nonlinear time-optimal MPC (TOMPC) seeks to steer nonlinear systems to a desired state in minimum time, thereby achieving highly performant control while accounting for system dynamics and constraints \cite{Verschueren2017Time, Florez2025practical}. Nevertheless, for nonlinear systems with fast dynamics, the high computational cost of online nonconvex optimization results in computation times often orders of magnitude higher than the available sampling interval \cite{dirckx2025asap}. A promising approach to overcome this limitation is to imitate the MPC control law with a neural network (NN), often through Behavioral Cloning (BC) \cite{bain2000aframework}, trained offline on MPC solutions \cite{karg2020efficient}. Given a sufficiently small NN that accurately reproduces the MPC control law, MPC-like control can be achieved at the high control frequencies required by fast dynamical systems. Such a controller is hereafter referred to as a neural network model predictive controller (NNMPC).

Existing literature already demonstrates real-time approximation of nonlinear MPC, but almost exclusively for quadratic-tracking objectives, and predominantly in simulation \cite{mohanty2025embedded, li2026neural}. Hardware demonstrations remain scarce, with existing works limited to either simpler objectives, soft or no terminal constraints or comparatively simple platforms \cite{tagliabue2024efficient, kong2025annmpc, hose2024parameter}. This work therefore tackles imitation of more computationally challenging MPC formulations, specifically a long-horizon, time-optimal formulation with hard terminal constraints. A key characteristic of such problems is that the computational cost of the MPC formulation not only restricts online deployment, but also strongly limits the generation of training data. Moreover, time-optimal control laws tend toward bang-bang behavior. Accurately approximating such laws with a smooth NN requires dense data near the switching points, increasing the demand for expert data. This makes conventional expert-data collection approaches such as Dataset Aggregation (DAgger) \cite{ross2010efficient} particularly costly, motivating more efficient expert-sampling strategies. Finally, existing works address individual challenges of NNMPC, such as dataset correlation \cite{castillo2025neural}, suboptimal expert labels \cite{zhang2021confidence}, controller adaptability  \cite{hose2024parameter}, constraint satisfaction \cite{mohanty2025embedded} and covariate shift \cite{ross2010efficient}, but a complete architecture integrating these solutions is currently lacking, leaving the design of a reliable NNMPC system fragmented across the literature.  
Therefore, this work aims at (i) presenting an expert-sampling algorithm, referred to as \emph{self-informed sampling} (SIFS), as a data-efficient alternative to DAgger specifically designed for time-optimal MPC, (ii) providing a complete training and control architecture that jointly addresses dataset correlation, suboptimal expert labels, controller adaptability, constraint satisfaction, covariate shift and efficient expert sampling, and (iii) providing real-world validation by deploying the resulting controller on a physical Furuta pendulum. 

\section{Prior Work} 
\label{sec:priorwork}
Many approaches have been proposed to overcome the computational limitations of MPC. One approach shifts the computational burden offline by storing precomputed optimal control solutions in a lookup table and interpolating them online, known as explicit MPC \cite{alberto2002explicit}. Neural networks offer a scalable alternative by learning a continuous mapping from states to control inputs. 
While NNMPC appears promising, several challenges remain.
\\\\ 
\noindent\textbf{Dataset Correlation.}  As argued in \cite{castillo2025neural}, MPC datasets typically violate the i.i.d. assumption underlying standard imitation learning \cite{ross2010efficient}, due to strong correlations both within and across trajectories. To mitigate this, \cite{castillo2025neural} proposes an Optimally-Sampled Dataset algorithm, subsampling the existing dataset so as to minimize biases that hinder generalization.
\\\\
\noindent\textbf{Suboptimality and Modality.} Nonconvex MPC can yield locally optimal or multimodal demonstrations. Consequently, different solutions may assign different actions to the same state, resulting in conflicting supervision. Under a mean squared error objective, these actions are effectively averaged, yielding a policy that follows neither solution and may therefore exhibit suboptimal behavior.  

Works such as \cite{zhang2025scizor} address this by filtering out suboptimal samples, yet works such as \cite{yang2021trail} argue that such samples can still provide valuable state-space coverage and should not be discarded outright. An alternative is weighted BC, which assigns each sample a data-dependent weight based on its estimated importance \cite{xu2023discriminator}. Such weights can steer the learned solution towards one mode (preferably the most optimal one), while simultaneously retaining state-space coverage. 

Nevertheless, to compute these weights, existing approaches require substantial additional training complexity, such as an auxiliary discriminator network \cite{xu2023discriminator} or bilevel optimization over ranked reference trajectories \cite{zhang2021confidence}. 
\\\\
\noindent\textbf{Data Efficiency.}   
\noindent Although imitating an MPC policy via BC is straightforward, covariate shift \cite{ross2010efficient} should be taken into account. Several strategies have been proposed to mitigate this issue, including augmenting the training data with perturbed states through noise injection \cite{laskey2017dart}, and DAgger \cite{ross2010efficient}. The latter iteratively executes the learned policy, queries the expert for actions at the visited states, and aggregates these newly labeled samples into the training set for subsequent retraining. 

Nevertheless, all such methods are sample-inefficient as they do not distinguish between states that require additional supervision and those already handled well by the policy \cite{cui2019uncertainty}. Since NN inference is computationally inexpensive compared to querying the expert, the learned policy can be used to identify states where its performance remains suboptimal. Expert demonstrations are then generated only for these states, focusing data collection on regions where additional supervision is most beneficial. This approach mitigates covariate shift, while simultaneously improving sample efficiency.
Previous work in data-efficient sampling employs epistemic uncertainty \cite{cui2019uncertainty} using ensembles \cite{lakshminarayanan2017simple} or Monte Carlo (MC) dropout \cite{gal2016dropout}. 
Similar to DAgger, these methods iteratively expand the training dataset, but query expert labels only for visited states with high epistemic uncertainty. A related approach, STRIDE (Sampling Trajectories for the Robust Imitation and Dynamic Evaluation) \cite{giusti2026stride}, avoids uncertainty estimation and instead samples states based on the distance between closed-loop NNMPC and MPC trajectories, combined with grid-based validation over the task space. Despite their different state-selection criteria, we argue that both epistemic uncertainty and STRIDE share three fundamental limitations. 

First, their selection metrics are indirect proxies for control quality: epistemic uncertainty can be overconfident \cite{lakshminarayanan2017simple}, while trajectory distance assumes that states far from expert trajectories are poorly controlled and nearby states are controlled well. Thus, queries may miss failure-prone states and waste samples on well-controlled ones. Second, neither metric accounts for the sensitivity of the system to control errors, treating equivalent errors identically despite potentially different consequences. Third, neither explicitly considers how control errors propagate through the system dynamics: an error that is readily corrected at the next timestep is treated the same as one that causes instability. Moreover, STRIDE requires evaluation against a grid whose size grows combinatorially with task-space dimension, limiting its scalability to higher-dimensional problems.  
\\\\
\noindent\textbf{Constraint Satisfaction.} Unlike classical MPC, naive neural imitation does not guarantee strict satisfaction of the originally imposed state and input constraints \cite{zieger2020towards}. Previous work proposes to enhance the neural training loss function with an additional penalty to include constraint violation into the learning gradient. This can be done via the penalty method (quadratically penalizing constraint violations), or logarithmic barrier methods (a loss term whose gradient grows unboundedly as the solution approaches the constraint boundaries) \cite{mohanty2025embedded, adhau2021constrained}. Notably, works employing such methods simultaneously train the network to learn the dual variables. Finally, Control Barrier Functions (CBFs) have been shown to be effective in guaranteeing constraint satisfaction when applied to the neural network output \cite{ames2017control}.
\\\\
\noindent\textbf{Adaptability.}
Unlike classical MPC, whose behavior can be adjusted online by retuning cost or model parameters, a trained NN policy offers no direct mechanism for tuning. Adapting expert parameters requires regenerating the dataset and retraining from scratch. Recent work \cite{hose2024parameter} mitigates this by learning an auxiliary NN alongside the nominal policy that predicts how the control output should change under parameter perturbations, enabling adaptation without retraining. Nevertheless, this requires substantial additional training effort, as accurate sensitivity estimates are crucial for performance but significantly increase the computational burden \cite{hose2024parameter}.

\section{Methodology} 
\begin{figure}[t]
    \centering
    \begin{tikzpicture}[
        node distance=0.6cm,
        font=\footnotesize,
        box/.style={draw, rounded corners, minimum width=1.2cm, minimum height=0.8cm, align=center, line width=1.0pt},
        bluebox/.style={box, fill=cyan!15, draw=cyan!60!black},
        sharpbox/.style={draw, minimum width=1.5cm, minimum height=0.8cm, align=center, line width=1.0pt},
        diamondnode/.style={draw, diamond, aspect=2, align=center, line width=1.0pt},
        circlenode/.style={draw, circle, minimum size=1.0cm, align=center, line width=1.0pt},
        arrow/.style={->, thick, >=stealth},
        legendbox/.style={draw, rounded corners=1pt, minimum width=0.4cm, minimum height=0.3cm, line width=0.8pt},
        parallelnode/.style={
            draw,
            shape=trapezium,
            trapezium left angle=80,
            trapezium right angle=100,
            minimum width=1cm,
            minimum height=1cm,
            align=center,
            text width=1.9cm,
            line width=1.0pt
        }
    ]
    
    
    \node[box, fill=cyan!12 , draw=cyan!60!black] (sampling)
    {
    \textbf{ \phantom{.}Optimal Sampling \phantom{.}}\\[-5pt]
    \rule{2.0cm}{0.01pt}\\[0pt]
    {\footnotesize Reduce data}\\
    {\footnotesize correlation}
    };
 
    \node[box, below=0.25cm of sampling, fill=cyan!12 , draw=cyan!60!black] (weight)
    {
    \textbf{Weighting} \\ \textbf{Suboptimal Data}\\[-5pt]
    \rule{2.0cm}{0.01pt}\\[0pt]
    {\footnotesize Resolve conflicting}\\{\footnotesize expert labels}
    };
 
    \node[box, below= 0.25cm of weight, fill=cyan!12 , draw=cyan!60!black] (train)
    {Train NN};
 
    \node[box, right=0.35cm of train, fill=blue!12, draw=blue!60!black] (nnmpc)
    {Final NNMPC};
    
    \node[box, below=0.25cm of train, fill=cyan!12 , draw=cyan!60!black] (rollouts) {Simulate \\Trajectories};
            
    \node[box, left=0.25cm of rollouts, fill=cyan!12 , draw=cyan!60!black] (informative)
    {
    \textbf{SIFS}\\[-5pt]
    \rule{2.0cm}{0.01pt}\\[0pt]
    {\footnotesize Minimize expert }\\{\footnotesize queries via }\\{\footnotesize targeted state }\\{\footnotesize selection }      
    };
 
    \node[box, above=0.25cm of informative, fill=red!10, draw=red!60!black, blur shadow={shadow blur steps=20}, line width=1.2pt] (expert)
    {Expert MPC};

    \node[box, above=0.25cm of expert, fill=cyan!12 , draw=cyan!60!black] (aggregation)
    {Dataset\\Aggregation};
    
    \node[box, left=0.25cm of aggregation, fill=cyan!12 , draw=cyan!60!black] (init)
    {Initial\\Dataset};
 
    \node[box, left=0.25cm of expert, fill=red!10, draw=red!60!black] (guess)
    {Smart\\Initial\\Guess};
    
    \node[box, below=0.25cm of nnmpc, fill=blue!12, draw=blue!60!black] (anc)
    {
    \textbf{Ancillary}\\ \textbf{Controller}\\[-5pt]
    \rule{1.5cm}{0.01pt}\\[0pt]
    {\footnotesize Boost robustness}\\{\footnotesize to plant--model}\\{\footnotesize mismatch}
    };
 
    \node[box, below=0.25cm of anc, fill=blue!12, draw=blue!60!black] (safety)
    {
    \textbf{Safety Filter}\\[-5pt]
    \rule{1.5cm}{0.01pt}\\[0pt]
    {\footnotesize Ensure constraint }\\{\footnotesize satisfaction}
    };
 
    \node[circlenode, left=0.4cm of safety, fill=green!10, draw=green!60!black] (plant)
    {
    \text{Plant}\\[-5pt]
    };
    
    \draw[arrow] (init) -- (aggregation);
    \draw[arrow] (sampling) -- (weight);
    \draw[arrow] (weight) -- (train);
    \draw[arrow] (train) -- (rollouts);
    
    \draw[arrow] (train) -- (nnmpc);
    \draw[arrow] (nnmpc) -- node[right, inner sep=2pt] {$u_{\mathrm{nom}}$} (anc);
    \draw[arrow] (anc) -- node[right, inner sep=2pt] {$u$} (safety);
    
    \draw[arrow] (safety) -- (plant);
    
    \draw[arrow] (rollouts) -- (informative);
    \draw[arrow] (informative) -- (expert);
    \draw[arrow] (expert) -- (aggregation);
    
    \draw[arrow] (guess) -- (expert);
    \draw[arrow] (expert) to[bend left=15] (guess);
    \draw[arrow] (aggregation) |- (sampling);
 
    \node[badge] at ([xshift=3pt,yshift=-3pt]expert.north west)      {I};
    \node[badge] at ([xshift=3pt,yshift=-3pt]aggregation.north west) {II};
    \node[badge] at ([xshift=3pt,yshift=-3pt]sampling.north west)    {III};
    \node[badge] at ([xshift=3pt,yshift=-3pt]weight.north west)      {IV};
    \node[badge] at ([xshift=3pt,yshift=-3pt]train.north west)       {V};
    \node[badge] at ([xshift=3pt,yshift=-3pt]rollouts.north west)    {VI};
    \node[badge] at ([xshift=3pt,yshift=-3pt]informative.north west) {VII};
    \node[badge] at ([xshift=3pt,yshift=-3pt]anc.north west)         {VIII};
    \node[badge] at ([xshift=3pt,yshift=-3pt]safety.north west)      {IX};
 
    \matrix [
        fill=white, 
        matrix of nodes, 
        nodes={anchor=west, inner sep=2pt}, 
        font=\footnotesize\bfseries, 
        column sep=5pt,
        row sep=1pt,
        anchor=south west,
        at={(current bounding box.south west)},
        xshift=0.cm, 
        yshift=-0.6cm
    ] {
        \node[legendbox, fill=red!10, draw=red!60!black] {}; & MPC \\
        \node[legendbox, fill=cyan!12, draw=cyan!60!black] {}; & NN Training Loop \\
        \node[legendbox, fill=blue!12, draw=blue!60!black] {}; & Controller \\
        \node[legendbox, fill=green!10, draw=green!60!black] {}; & Hardware \\
    };
    \end{tikzpicture}
    \caption{Schematic outline of the proposed integrated architecture. Roman numerals refer to the blocks as introduced in the text.}
    \label{fig:thesis_overview}
\end{figure}
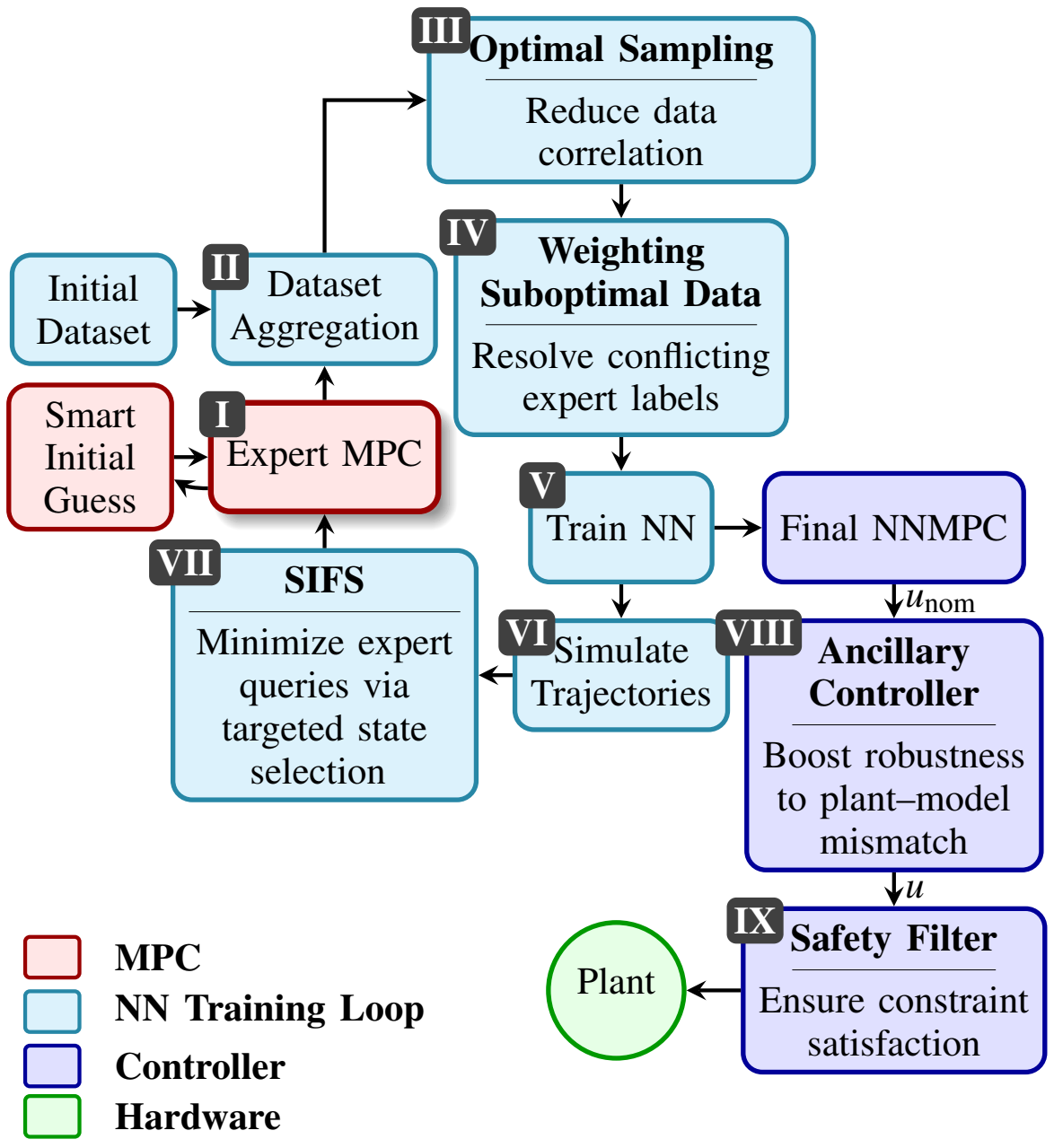

This work expands on prior work by providing a complete training
and control architecture that jointly addresses the challenges
of NNMPC, discussed in the previous section. A schematic representation of this architecture is provided in Fig.~\ref{fig:thesis_overview}. First, an expert MPC\blk{I} is used to generate new data points, which are iteratively aggregated to the dataset\blk{II}. The resulting dataset is then filtered using an optimal sampling algorithm\blk{III} to remove correlated and redundant samples. Each remaining data point is assigned a weight\blk{IV} to reduce the influence of conflicting expert labels, and the NN is subsequently trained\blk{V} on the weighted dataset using BC. The current policy is evaluated through closed-loop rollouts\blk{VI}, with the proposed SIFS algorithm\blk{VII} identifying poorly performing states. The expert MPC is queried only on these states. The resulting expert demonstrations are added to the dataset, completing the data-aggregation loop. The NNMPC output is adjusted by an ancillary controller\blk{VIII} for plant--model mismatch compensation and a safety filter\blk{IX} for constraint satisfaction. The following sections detail each component.

\subsection{Expert Model Predictive Controller}
The expert MPC considered here involves a long-horizon, nonconvex, time-optimal formulation with a hard terminal constraint, where time-optimality is realized using an exponentially weighted objective following \cite{Florez2025practical}. Unlike quadratic-tracking MPC, it exploits the full actuation range to reach the target in minimum time without requiring a reference trajectory. Nevertheless, the considered time-optimal formulation is sensitive to initialization and computationally demanding.

For TOMPC, the time-to-go (T2G) naturally serves as a value function, defined as the first time the predicted MPC trajectory from an initial state $\mathbf{x}_0$ lies within a tolerance $\xi$ of the reference $\mathbf{x}_f$:
$
    \mathrm{T2G}(\mathbf{x}_0) = \Delta t \cdot \min\{k : \|\mathbf{x}_k - \mathbf{x}_f\| \le \xi\},
$
where $\Delta t$ is the sampling time and $\mathbf{x}_k$ is the predicted state at step $k$ of the MPC solution.

\subsection{Correlation between expert labels} \label{subsec:correlation}
Since expert data generated by an MPC are strongly correlated, raw MPC datasets often oversample regions containing little variation in the optimal control law \cite{castillo2025neural}. This adds redundancy to the dataset and simultaneously skews the cost minimization, making the overall trained policy worse. Therefore, our proposed framework builds upon the approach of \cite{castillo2025neural}, in which the NN is trained on an Optimally-Sampled Dataset, preserving the information content of the MPC policy at a prescribed resolution while eliminating redundant data points.

\subsection{Conflicting expert labels}  \label{subsec:conflicting}
Since the nonconvex optimal control problem may converge to different local optima, nearby states can receive conflicting expert labels. To mitigate this, this work employs weighted BC, multiplying each quadratic error term in the training loss by a weight, thereby giving higher-quality samples greater influence during training.

To efficiently compute weights, we propose a lightweight T2G-based weighting strategy, leveraging T2G as a direct measure of trajectory optimality under a time-optimal objective. Concretely, we assign each training sample a weight $w_\mathrm{knn}(\mathbf{x}_i)$ based on its T2G relative to the distance-weighted average T2G of its $M$ nearest neighbors, excluding samples from the same trajectory:
\begin{equation}
\alpha_{ij} = e^{-\frac{\|\mathbf{x}_i - \mathbf{x}_j\|^2}{\sigma^2}}, \quad \mathrm{\overline{T2G}}(\mathbf{x}_i) =
\frac{\sum\limits_{j \in \mathcal{N}_M(i)} \alpha_{ij} \, \mathrm{T2G}(\mathbf{x}_j)}
     {\sum\limits_{j \in \mathcal{N}_M(i)} \alpha_{ij}},
\end{equation} 

\begin{equation}
w_\mathrm{knn}(\mathbf{x}_i) = \frac{1}{1 + e^{p\left(\mathrm{T2G}(\mathbf{x}_i) - \mathrm{\overline{T2G}}(\mathbf{x}_i)\right)}} \cdot q + (1 - q).
\end{equation}
\noindent Here, $\alpha_{ij}$ is a distance-based Gaussian weight, $\mathcal{N}_M(i)$ denotes the $M$ nearest neighbors of $\mathbf{x}_i$, excluding points from its own trajectory; $\sigma$ controls the influence of distance in state space, while $p\in \mathbb{R}^+$ and $q \in [0,1]$ control the sigmoid sharpness and lower bound, respectively. By incorporating distance-based weighting in the computation of $\mathrm{\overline{T2G}}$, trajectory points in sparsely sampled regions retain higher influence, even when they appear suboptimal relative to their neighbors. This balances state-space coverage with the suppression of suboptimal expert data, although conflicting labels from equally time-optimal modes remain unresolved.

\subsection{Self-Informed Sampling} \label{sec:informative} 
We introduce self-informed sampling as a data-efficient sampling strategy that addresses the limitations of existing approaches discussed in Section~\ref{sec:priorwork}. Concretely, we augment the NN with an additional output head that predicts a value function, more specifically the T2G. The T2G decreases by one sampling interval per time step along a time-optimal trajectory, providing a direct indicator of deviations from optimal behavior without requiring the expert controller.

\begin{figure}[t]
    \centering
    \includegraphics[width=1.0\columnwidth]{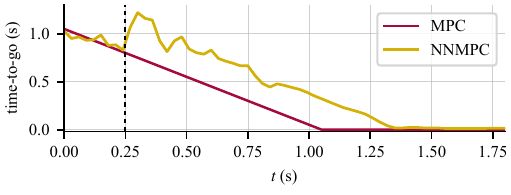}
    \caption{Predicted time-to-go (T2G) along a rollout for the expert MPC (red) and a partially trained NNMPC (yellow). While the NNMPC prediction is noisy, it closely follows the expected trend except for the highlighted transition (dashed line), where it jumps to a different trajectory. This deviation indicates suboptimal behavior and identifies the states most informative for expert querying.}
    \label{fig:t2gillustration}
\end{figure}

Since the T2G is learned, its predictions contain noise and occasional harmless discontinuities. The informative-state criterion should therefore tolerate short-term deviations and isolated jumps as long as the overall decreasing trend remains consistent. Conversely, persistent deviations, such as stagnation, gradual increases, or strongly irregular behavior, should be detected and flagged from their onset. To this end, we design a simple algorithm to identify segments that follow the expected linear T2G evolution and flag transitions between them (see Fig.~\ref{fig:t2gillustration}). Since the exact state causing the deviation may be missed, states preceding a flagged state are also considered as candidates. Candidates are then sampled for expert querying with a probability that increases with (i) their proximity to a flag, (ii) how early they occur in the rollout, as early errors propagate through the remainder of the trajectory, and (iii) the \emph{severity} of the corresponding rollout,  which quantifies how much the predicted T2G deviates from the expected decrease of one sampling interval per time step. 

Unlike epistemic-uncertainty or trajectory-distance criteria, which judge the policy at a single state through an indirect proxy, the proposed criterion evaluates the T2G over an entire rollout. Because the network receives only the current state as input, with no access to its previous T2G estimate, it cannot produce a consistent T2G sequence by simply decrementing an earlier output. A T2G prediction that decreases by exactly one sampling interval per step is therefore unlikely to occur by chance, and this likelihood decreases rapidly with the rollout length. Such a trend is unlikely to arise unless the network has (i) learned the T2G of the visited states and (ii) predicts controls that steer the system to the correct next state. This avoids overconfidence in poorly handled states, as well as unnecessary queries at states that are already well learned. Furthermore, locally suboptimal actions with negligible effect on the trajectory leave the trend intact and are not flagged, whereas errors that propagate through the dynamics break it.

\subsection{Controller Adaptability} \label{subsec:anc} 
To compensate for plant--model mismatch without retraining on real-world data, the NNMPC is augmented with a lightweight ancillary controller that corrects its control action online. Two options are considered: a dynamic-inversion compensator (DIC) and a Linear Quadratic Regulator (LQR).
\\

\subsubsection{Dynamic-Inversion Compensator}
Following the inverse-dynamics transfer approach of \cite{christiano2016transfer}, the DIC compensates for plant--model mismatch by computing an input $\bar{u}$ that reproduces, on an estimated plant model, the output dynamics induced by the nominal NNMPC input $u_{\mathrm{nom}}$ on the nominal model. Unlike learned inverse models, the DIC analytically inverts the plant dynamics, similar to input--output feedback linearization \cite{Slotine1991AppliedNC}.

Consider a single-input single-output (SISO) control-affine system
$\dot{\mathbf{x}} = f(\mathbf{x};\boldsymbol{\theta}) + g(\mathbf{x};\boldsymbol{\theta})\,u$
with parameters $\boldsymbol{\theta}$ and output $y = h(\mathbf{x})$. Differentiating the output $r$ (relative degree) times until the input explicitly appears yields
\begin{equation}
y^{(r)} = a(\mathbf{x};\boldsymbol{\theta}) + b(\mathbf{x};\boldsymbol{\theta})\,u,
\qquad
a = L_f^{r} h,\;\; b = L_g L_f^{r-1} h,
\label{eq:affine}
\end{equation}
with $L_f$ and $L_g$ the Lie derivatives along $f$ and $g$. 

At each control step, the nominal model (with model parameters $\boldsymbol{\theta}_{\mathrm{n}}$) is first evaluated using $u_{\mathrm{nom}}$ to determine the desired output derivative:
$
v = a(\mathbf{x};\boldsymbol{\theta}_{\mathrm{n}})
  + b(\mathbf{x};\boldsymbol{\theta}_{\mathrm{n}})\,u_{\mathrm{nom}}.
$
The DIC then analytically inverts the estimated plant dynamics (with $\hat{\boldsymbol{\theta}}$ the estimate of the true plant parameters) to find the input $\bar{u}$ that reproduces this derivative on the real plant:
\begin{equation}
\bar{u} =
\frac{v - a(\mathbf{x};\hat{\boldsymbol{\theta}})}
     {b(\mathbf{x};\hat{\boldsymbol{\theta}})},
\qquad
b(\mathbf{x};\hat{\boldsymbol{\theta}}) \neq 0.
\label{eq:dic}
\end{equation}

For multiple-input, multiple-output (MIMO) systems, exact matching of only some output dynamics may conflict with the overall NNMPC objective (also regulating other output dynamics). Therefore, the applied input is obtained by blending the nominal and dynamically compensated inputs:
$
u = u_{\mathrm{nom}} + \zeta\,(\bar{u} - u_{\mathrm{nom}}),
$
with $\zeta \in [0,1]$. 

The DIC thus requires a control-affine model with a well-defined relative degree and a nonzero input coefficient $b$ throughout the operating region, as well as an estimate $\hat{\boldsymbol{\theta}}$ of the plant parameters.
\\
\subsubsection{ Linear Quadratic Regulator}
A more generic alternative employs an ancillary LQR, inspired by its use in the tube-based MPC framework of \cite{kim2025bridging}. At each time step, the state predicted by the nominal model given the previous state and input, $\bar{\mathbf{x}}$, is compared with the measured next state $\mathbf{x}$ of the real plant. The LQR then directly converts this observed mismatch into a corrective input:
\begin{equation}
    u = u_{\mathrm{nom}} + \mathbf{K}(\bar{\mathbf{x}} - \mathbf{x}),
    \label{eq:lqr}
\end{equation}
\noindent where $\mathbf{K}$ is the LQR gain, computed using the linearized nominal model. 
\subsection{Constraint Satisfaction} \label{subsec:safety} 
We address constraint satisfaction at two levels: additional terms in the training loss encourage the neural network to learn the constraints imposed by the original MPC formulation, while a differentiable safety filter in the final network layer formally guarantees constraint satisfaction.

To steer the raw network output towards the feasible region, the training loss combines a logarithmic barrier with a quadratic penalty:
\begin{align}
\label{eq:barrier}
L_{\mathrm{barrier}}
&= -\operatorname{mean}\Big[
    \log(u_{\mathrm{raw}}-u_{\min,\mathrm{safe}}+\epsilon)
    \nonumber\\[-1mm]
&\qquad\qquad
    +\log(u_{\max,\mathrm{safe}}-u_{\mathrm{raw}}+\epsilon)
    \Big], \\
\label{eq:penalty}
L_{\mathrm{penalty}}
&= \operatorname{mean}\Big[
    \max(0,u_{\mathrm{raw}}-u_{\max,\mathrm{safe}})^2
    \nonumber\\[-1mm]
&\qquad\qquad
    +\max(0,u_{\min,\mathrm{safe}}-u_{\mathrm{raw}})^2
    \Big].
\end{align}
Here, $u_{\mathrm{raw}}$ denotes the neural network output before the safety filter, $u_\mathrm{min/max,safe}$ denote the corresponding safe bounds imposed by the safety filter and $\epsilon$ ensures numerical stability. The barrier term is evaluated only within the safe region; outside it, its contribution is zero.

The network is not trained to jointly predict the dual variables \cite{mohanty2025embedded, adhau2021constrained}, as this would complicate data generation. Without a dual-variable term, the logarithmic barrier \eqref{eq:barrier} alone proved insufficient in our experiments: it provides no gradient when the raw output violates the safe bounds, leading to unstable training. The quadratic penalty \eqref{eq:penalty} complements the barrier by providing a gradient in the infeasible region, thereby mitigating this instability.

The safety filter projects the raw network output onto the set of inputs that keep the system within its constraints at all future times: $u_{\mathrm{safe}} = \arg\min_{u \in \mathcal{U}_{\mathrm{s}}(\mathbf{x})} \|u - u_{\mathrm{raw}}\|^2$, with $\mathcal{U}_{\mathrm{s}}(\mathbf{x}) = \{u \in \mathcal{U} : f_d(\mathbf{x}, u) \in \mathcal{C}\}$, $\mathcal{C}$ a control-invariant (safe) set and $f_d(\mathbf{x}, u)$ the discrete system dynamics. Ideally, $\mathcal{C}$ is the maximal control-invariant set, making $\mathcal{U}_{\mathrm{s}}(\mathbf{x})$ maximally permissive and the filter nonconservative, without additional constraints on the MPC \cite{hsu2024safety}. If this set is intractable, CBFs \cite{ames2017control} provide a conservative alternative. The same CBF constraint should then be included in the expert MPC, preventing the NNMPC from learning infeasible control actions.

\section{Experimental Setup}
\subsection{Furuta Pendulum and Expert Controller}
To demonstrate the capability of the proposed framework, we consider the imitation of a TOMPC for control of a Furuta pendulum. The state vector and control input are defined as $\mathbf{x} = (\theta_1, \theta_2, \dot{\theta}_1, \dot{\theta}_2)$ and $u = \ddot{\theta}_1$, with $\theta_1$ and $\theta_2$ denoting the base and pendulum-arm angles, respectively, and $\theta_2 = \pi$ rad corresponding to the inverted position. The continuous-time nonlinear dynamics $\dot{\mathbf{x}} = f(\mathbf{x}) + g(\mathbf{x}) \cdot u$ follow \cite{Cazzolato2011Furuta}. The TOMPC formulation itself is:

\begin{equation} \label{eq:OCP_Furuta}
    \begin{aligned}
        \min_{\mathbf{x}, \mathbf{u}, s} \quad & \sum_{k=0}^{N-1} \left[ \left(\gamma^{\frac{\kappa \cdot k}{N}}\right) \cdot \| \mathbf{x}_k - \mathbf{x}_f \|_2^2 + w_u \cdot u_k^2 + w_s \cdot s^2 \right] \\
        \text{subject to:} \quad & \mathbf{x}_{k+1} = f_d(\mathbf{x}_k, u_k) \\
        & \mathbf{x}_0 = \mathbf{x}_\mathrm{init}, \quad \mathbf{x}_N = \mathbf{x}_f \\
        & \theta_\mathrm{1,min} \leq \theta_{1, k} \leq \theta_\mathrm{1, max} , \hspace{7px} \qquad \forall k \in \{0, \dots, N\} \\
        & u_\mathrm{min} \leq u_k \leq u_\mathrm{max} , \hspace{5px} \qquad \qquad \forall k \in \{0, \dots, N-2\} \\
        & -s \leq u_{N-1} \leq s, \quad  s \geq 0.    
    \end{aligned}
\end{equation}

\noindent The MPC predicts over a finite horizon of $N = 60$ steps with sampling time $\Delta t = 0.03$~s. The state $\mathbf{x}_k\in\mathbb{R}^4$ and control input $u_k\in\mathbb{R}$ are optimized toward the target state $\mathbf{x}_f\in\mathbb{R}^4$, subject to the input bounds $u_{\min}\leq u_k\leq u_{\max}$ and arm-angle bounds $\theta_{1,\min}\leq\theta_{1,k}\leq\theta_{1,\max}$. The objective uses $w_u = 1$ to penalize control effort and $w_s = 10^5$ to penalize the slack variable $s\in\mathbb{R}$, while $\gamma = 150$ denotes the base weight and $\kappa =3.0$ the exponent of the time-dependent weighting. The discrete-time transition $\mathbf{x}_{k+1} = f_d(\mathbf{x}_k, u_k)$ is obtained via fourth-order Runge-Kutta (RK4) integration of the continuous-time dynamics using timestep $\Delta t$.

The MPC is formulated to swing up and stabilize the Furuta pendulum within a single prediction horizon, while respecting box constraints on the base angle and control input. The objective is to achieve this task in minimum time while converging the base angle to its nominal upright position. Time-optimality is implemented using an exponential weight \cite{Verschueren2017Time, Florez2025practical}. A vanishing terminal control input is used to encourage steady-state behavior upon reaching the target terminal state. 
The optimal control problem is transcribed using multiple shooting with Fatrop~\cite{vanroye2023fatrop}, within the Impact library \cite{Florez2023impact}, leveraging its C-API for faster execution.

Empirically, a good initial guess is crucial for solver convergence. Following \cite{Lembono2020memory}, we therefore maintain a growing database of previously computed feasible trajectories. A KD-tree queries this database for nearby trajectories, which provide warm starts for subsequent optimizations.

\subsection{NNMPC Training Setup}

The NN is a Multilayer Perceptron (MLP), with three fully connected hidden layers of 512 nodes each and $\tanh$ activations, mapping the current state $\mathbf{x}_k$ and target state $\mathbf{x}_f$ (eight inputs) to the T2G and the nominal control output ${u}_{k, \mathrm{raw}}$ (two outputs).
The latter is subsequently passed through the safety layer to obtain the applied control input $u_{k, \mathrm{safe}}$. 
The multi-objective loss is formulated as:
\begin{equation}
L_\mathrm{total} = \lambda_u L_{u} + \lambda_\mathrm{T2G} L_\mathrm{T2G} + \lambda_{b}L_\mathrm{barrier} + \lambda_{p} L_\mathrm{penalty},
\end{equation}
where each term $L_i$ is computed as a weighted mean $\frac{\sum w_\mathrm{knn} \cdot \ell_i}{\sum w_\mathrm{knn}}$, with $w_\mathrm{knn}$ the $k$-nearest neighbors weights of Section~\ref{subsec:conflicting}, and $\lambda_i$ hyperparameters. The imitation error functions employed are: $\ell_u = (u_\mathrm{safe} - u_\mathrm{expert})^2, \quad 
    \ell_{\mathrm{T2G}} = |\mathrm{T2G} - \mathrm{T2G}_\mathrm{expert}|.$
The barrier and penalty terms are defined in Section~\ref{subsec:safety}.

\subsection{Training Procedure}
First, the network is pre-trained on an initial grid-based dataset of 15,601 data points for 1200 epochs, using a learning rate of $8\times10^{-4}$.
Then, the iterative training procedure presented in Section~\ref{sec:informative} is run for 1441 iterations, aggregating 1,189,320 data points, of which 708,312 remained after optimal sampling (Section~\ref{subsec:correlation}). Training is done using minibatches of 4096 samples, with 70 epochs per iteration and at most 15 new trajectories (900 data points) added per iteration with a fixed learning rate of $10^{-6}$. 
 
\section{Closed-loop Validation}
\subsection{Simulation}
\begin{figure}[t]
    \centering
    \subfloat[Distribution of convergence times for the NNMPC and the expert MPC.\label{fig:swimlanes_dist}]{%
        \includegraphics{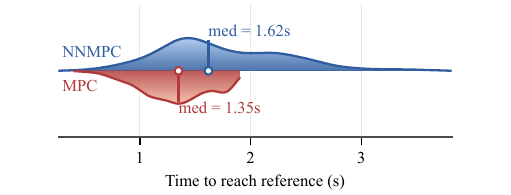}}\\
    \subfloat[Distribution of pairwise convergence-time differences.\label{fig:swimlanes_diff}]{%
        \includegraphics{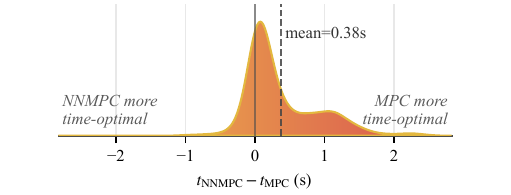}}
    \caption{Performance comparison of the expert MPC and NNMPC over 1000 randomly sampled initial states.}
    \label{fig:swimlanes}
\end{figure}

Fig.~\ref{fig:swimlanes} compares the convergence performance of the expert MPC and NNMPC over 1000 randomly sampled initial states within the region of interest. Each state is evaluated independently with both controllers over a 3.6-second evaluation window, twice the MPC prediction horizon. Both controllers successfully converge in $91.5\%$ of the cases, while only the NNMPC succeeds in $4.2\%$ and only the MPC in $4.3\%$.

Fig.~\ref{fig:swimlanes_dist} shows that the NNMPC takes slightly longer to reach the reference state, with a median time of 1.62~s compared to 1.35~s for the MPC. However, Fig.~\ref{fig:swimlanes_diff} reveals a nonnegligible number of cases in which the NNMPC reaches the reference faster than the expert MPC. This indicates that the expert MPC is not always time-optimal and that the NNMPC can generalize beyond the demonstrated trajectories, in some cases achieving faster convergence.
The remaining difference in convergence time may result from insufficient training or suboptimal handling of conflicting labels in the MPC-generated dataset.

\subsection{Computation time} 
Table~\ref{tab:timing} compares the per-iteration computation time of the original MPC with the proposed NNMPC executed in real time using CPU (Intel Core i7-9850H) and GPU (Nvidia Quadro P1000 Mobile) inference.
\begin{table}[b]
\centering
\caption{Per-iteration computation time during pendulum swing-up and the resulting maximum reliable control frequency.}
\label{tab:timing}
\renewcommand{\arraystretch}{1.15}
\setlength{\tabcolsep}{5pt}
\begin{tabular}{lcccc}
\toprule
 & \multicolumn{3}{c}{Computation time (ms)} & Max. reliable \\
\cmidrule(lr){2-4}
 & Mean & Std. dev. & Max & freq. (Hz) \\
\midrule
MPC         & 114.165 & 150.132 & 732.676 & 1.36 \\
NNMPC (CPU) & \textbf{4.839} & \textbf{0.809} & \textbf{10.576} & \textbf{94.6} \\
NNMPC (GPU) & \textbf{0.409} & \textbf{0.094} & \textbf{0.998}  & \textbf{1002} \\
\bottomrule
\end{tabular}
\end{table}
When deployed on CPU, the NNMPC controller supports a maximum control frequency of $\approx$100~Hz. On GPU, inference consistently remains below 1~ms, theoretically enabling control frequencies up to 1~kHz. This represents a substantial improvement compared to the original MPC, whose maximum reliable control frequency of approximately 1.36~Hz severely limits its use for real-time control.

\section{Component Validation}
The individual building blocks of the proposed NNMPC architecture are evaluated through simulation. Closed-loop performance is quantified by randomly sampling initial states from a region of interest and recording whether the controller successfully reaches the terminal state within 3.6 seconds. The percentage of trajectories that fail to reach the terminal state within this time is referred to as the \textit{failed trajectory percentage}.

\subsection{Conflicting expert labels}
\begin{figure}[b]
    \centering
    \includegraphics[width=1.0\columnwidth]{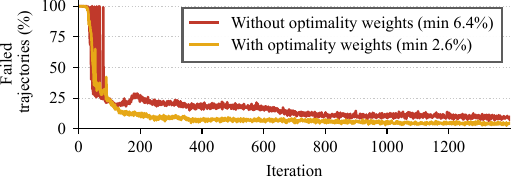}
        \caption{Evolution of the failed trajectory percentage over training iterations, with and without the proposed weighting strategy.}
    \label{fig:weights_ablation}
\end{figure}

Fig.~\ref{fig:weights_ablation} displays the results of ablating the proposed T2G-based weighting strategy of Section~\ref{subsec:conflicting}, showing that its removal leads to a nonnegligible increase in overall failed-trajectory percentage.

\subsection{Self-Informed Sampling}
\begin{figure}[t]
    \centering
    \subfloat[Failure rate over training iterations.\label{fig:sifs_failure}]{%
        \includegraphics{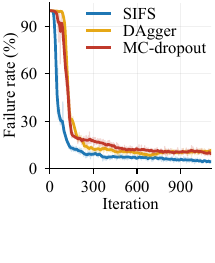}}\hfill
    \subfloat[Additional expert queries and training time needed to match the SIFS failure rate.\label{fig:sifs_gap}]{%
        \includegraphics{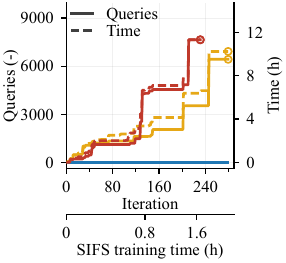}}
    \caption{Training progress of self-informed sampling (SIFS), DAgger, and MC-dropout.}
    \label{fig:ifs_comparison}
\end{figure}

Fig.~\ref{fig:ifs_comparison} evaluates the proposed self-informed sampling (SIFS) strategy by replacing the state-selection component of Section~\ref{sec:informative} with DAgger and MC-dropout (implemented according to \cite{cui2019uncertainty}). DAgger samples uniformly, and MC-dropout uses epistemic uncertainty estimation based on 20 forward passes with a dropout probability of 30\%. 
As shown, SIFS achieves a notably faster reduction in failure rate. Fig.~\ref{fig:sifs_gap} demonstrates that both DAgger and MC-dropout become increasingly less data- and time-efficient compared to SIFS. For this particular example, comparable performance is reached approximately three times faster and with less than half the required training data.

\subsection{Controller Adaptability}
\begin{figure}[b]
    \centering
    \includegraphics[width=1.0\columnwidth]{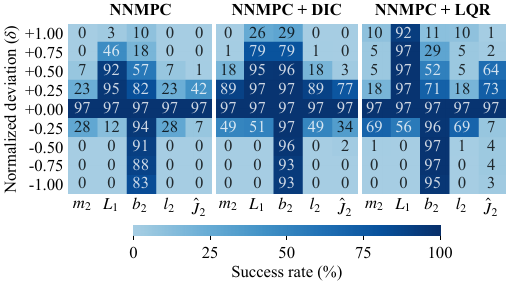}
    \caption{Each model parameter ($m_2$, $L_1$, $b_2$, $l_2$, $\hat{J}_2$; notation as in \cite{Cazzolato2011Furuta}) was systematically perturbed via $p \leftarrow p \cdot \left(1 + \delta/2\right)$, where $\delta$ denotes the normalized deviation, varying between $-1$ and $1$. For every parameter setting, 200 rollouts were performed from randomly sampled initial states. Each cell reports the percentage of successful trajectories.}
    \label{fig:anc_improv}
\end{figure}

Both the nominal and ancillary controllers were evaluated on systematically perturbed plant models. Fig.~\ref{fig:anc_improv} shows that both ancillary architectures improve robustness overall, with LQR performing better for some perturbations and DIC for others. The best choice therefore depends on the plant and the expected model mismatch. Rather than advocating a particular ancillary architecture, these results demonstrate that simple ancillary controllers can provide useful adaptability to model mismatch without retraining.

\section{Hardware Validation}
Finally, the NNMPC is deployed on a physical Furuta pendulum, using the C++ deployment framework ImpaC++t of \cite{bos2026impact}. The base is driven by a stepper motor, and the pendulum angle is measured by an incremental encoder (4096 ticks/rev). Both the motor controller and the encoder interface are connected via CANopen over EtherCAT, using the SOEM master library. The I/O thread runs at 1~kHz and the NNMPC at 250~Hz. The full pipeline runs on a laptop with standard Linux using a GPU for the network evaluation.

\begin{figure}[t!] 
\centering
\includegraphics[width=1.0\columnwidth]{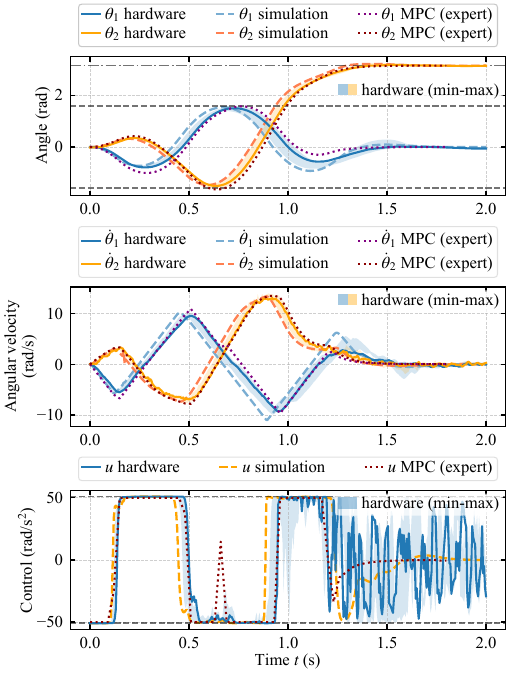}
\caption{State and control trajectories of the expert MPC solution, the NNMPC in simulation, and the NNMPC deployed on the physical pendulum. The solid hardware curves show one representative run. The shaded bands show the minimum-to-maximum envelope over four additional hardware reruns.} 
\label{fig:hardware_overlays}
\end{figure}

\begin{figure}[b]
    \centering
    \includegraphics[width=1.0\columnwidth]{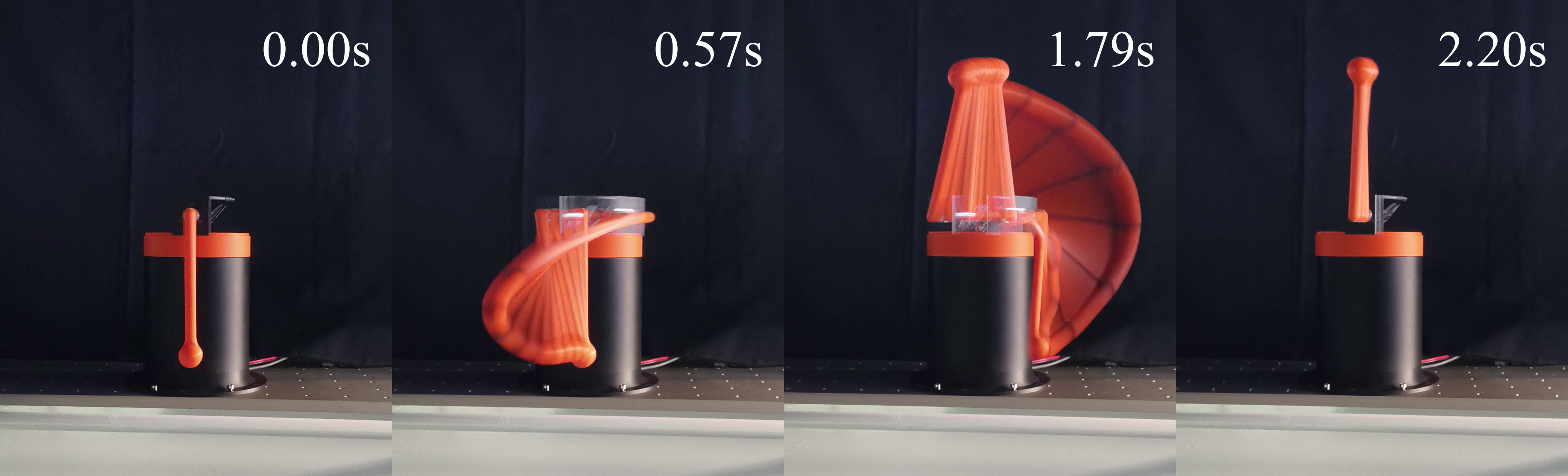}
    \caption{Swing-up and stabilization starting from the downward equilibrium. Timestamps correspond to the final frame in each overlaid sequence.}
    \label{fig:fivepanelres}
\end{figure}

The controller performance for the physical pendulum was assessed through a swing-up task. Fig.~\ref{fig:hardware_overlays} shows that the real-world state trajectory closely matches the simulated results. The main discrepancy is a high-frequency oscillatory control effort near the inverted equilibrium, attributable to sensor noise and residual plant--model mismatch. This did not compromise stability and could be further suppressed with a low-pass filter on the NNMPC output.
Fig.~\ref{fig:fivepanelres} displays the swing-up task on the physical setup. 




\section{Conclusion}
We presented an integrated architecture for imitating TOMPC with a neural network, which combines state-of-the-art techniques for dataset correlation, constraint satisfaction, and covariate shift with three original contributions tailored to TOMPC: self-informed sampling using a jointly learned T2G, a lightweight T2G-based weighting of conflicting expert labels, and ancillary controllers that compensate for plant--model mismatch without retraining.

On a Furuta pendulum, self-informed sampling matched the performance of state-of-the-art methods with less than half the training data and roughly one third of the training time, while the T2G-based weighting markedly reduced the failure rate. The resulting NNMPC attains a region of attraction on par with the expert MPC at an over 700-fold lower worst-case computation time (on GPU), and the ancillary controllers substantially improved generalization to perturbed plant models in simulation. On hardware, running at 250~Hz, the controller swings up and stabilizes the pendulum along trajectories closely matching simulation.

The remaining performance gap may stem from insufficient training or conflicting expert labels. Future work could explore multimodal architectures, such as Mixture Density Networks \cite{Morimasa2023approximation}, and continual learning to limit training-time growth as the dataset expands.

\addtolength{\textheight}{-0.1cm}   




\urlstyle{same}
\section*{APPENDIX}
Videos demonstrating the experimental results and further implementation details are available on the project webpage:
\url{https://maartendebaecke.github.io/PW/neural-tompc/}.

\section*{ACKNOWLEDGMENT}
The authors thank Alvaro Javier Florez Martinez for his support with the MPC design and the IMPACT software and Mathias Bos for his support with the hardware validation.


\bibliographystyle{IEEEtran}
\bibliography{references}

@IEEEtranBSTCTL{IEEEexample:BSTcontrol,
  CTLuse_forced_etal       = "yes",
  CTLmax_names_forced_etal = "3",
  CTLnames_show_etal       = "3",
  CTLuse_url               = "no"
}

@article{dirckx2025asap,
    author = {Dirckx, Dries and Bos, Mathias and Vandewal, Bastiaan and Vanroye, Lander and Swevers, Jan and Decr\'{e}, Wilm},
    title = {ASAP-MPC: an asynchronous update scheme for online motion planning with nonlinear model predictive control},
    year = {2025},
    issue_date = {Mar 2025},
    publisher = {Kluwer Academic Publishers},
    address = {USA},
    volume = {49},
    number = {1},
    issn = {0929-5593},
    url = {https://doi.org/10.1007/s10514-025-10192-w},
    doi = {10.1007/s10514-025-10192-w},
    journal = {Auton. Robots},
    month = mar,
    numpages = {17}
}

@ARTICLE{castillo2025neural,
  author={Castillo, Alberto and Pryor, Elliot and El Fathi, Anas and Kovatchev, Boris and Breton, Marc},
  journal={IEEE Transactions on Cybernetics}, 
  title={Neural Networks for On-Chip Model Predictive Control: A Method to Build Optimized Training Datasets and its Application to Type-1 Diabetes}, 
  year={2025},
  volume={55},
  number={7},
  pages={3062-3072},
  doi={10.1109/TCYB.2025.3562032}
}

@inproceedings{cui2019uncertainty,
    author = {Cui, Yuchen and Isele, David and Niekum, Scott and Fujimura, Kikuo},
    title = {Uncertainty-Aware Data Aggregation for Deep Imitation Learning},
    year = {2019},
    publisher = {IEEE Press},
    url = {https://doi.org/10.1109/ICRA.2019.8794025},
    doi = {10.1109/ICRA.2019.8794025},
    booktitle = {2019 International Conference on Robotics and Automation (ICRA)},
    pages = {761–767},
    numpages = {7},
    location = {Montreal, QC, Canada}
}

@misc{zhang2025scizor,
  title        = {SCIZOR: A Self-Supervised Approach to Data Curation for Large-Scale Imitation Learning},
  author       = {Zhang, Yu and Xie, Yuqi and Liu, Huihan and Shah, Rutav and Wan, Michael and Fan, Linxi and Zhu, Yuke},
  year         = {2025},
  eprint       = {2505.22626},
  archivePrefix= {arXiv},
  primaryClass = {cs.RO},
  url          = {https://arxiv.org/abs/2505.22626},
  note         = {Preprint},
howpublished = {arXiv:2505.22626}
}

@inproceedings{zhang2021confidence,
 author = {Zhang, Songyuan and Cao, ZHANGJIE and Sadigh, Dorsa and Sui, Yanan},
 booktitle = {Advances in Neural Information Processing Systems},
 editor = {M. Ranzato and A. Beygelzimer and Y. Dauphin and P.S. Liang and J. Wortman Vaughan},
 pages = {12340--12350},
 publisher = {Curran Associates, Inc.},
 title = {Confidence-Aware Imitation Learning from Demonstrations with Varying Optimality},
 url = {https://proceedings.neurips.cc/paper_files/paper/2021/file/670e8a43b246801ca1eaca97b3e19189-Paper.pdf},
 volume = {34},
 year = {2021}
}

@InProceedings{xu2023discriminator,
  title = 	 {Discriminator-Weighted Offline Imitation Learning from Suboptimal Demonstrations},
  author =       {Xu, Haoran and Zhan, Xianyuan and Yin, Honglei and Qin, Huiling},
  booktitle = 	 {Proceedings of the 39th International Conference on Machine Learning},
  pages = 	 {24725--24742},
  year = 	 {2022},
  editor = 	 {Chaudhuri, Kamalika and Jegelka, Stefanie and Song, Le and Szepesvari, Csaba and Niu, Gang and Sabato, Sivan},
  volume = 	 {162},
  series = 	 {Proceedings of Machine Learning Research},
  month = 	 {17--23 Jul},
  publisher =    {PMLR},
  url = 	 {https://proceedings.mlr.press/v162/xu22l.html},
}

@inproceedings{
    yang2021trail,
    title={{TRAIL}: Near-Optimal Imitation Learning with Suboptimal Data},
    author={Mengjiao Yang and Sergey Levine and Ofir Nachum},
    booktitle={International Conference on Learning Representations},
    year={2022},
    url={https://openreview.net/forum?id=6q_2b6u0BnJ}
}

@article{Cazzolato2011Furuta,
author = {Cazzolato, Ben and Prime, Zebb},
year = {2011},
month = {04},
pages = {},
title = {On the Dynamics of the Furuta Pendulum},
volume = {2011},
journal = {Journal of Control Science and Engineering},
doi = {10.1155/2011/528341}
}

@InProceedings{hose2024parameter,
  title = 	 {Parameter-adaptive approximate {MPC}: {T}uning neural-network controllers without retraining},
  author =       {Hose, Henrik and Gr\"{a}fe, Alexander and Trimpe, Sebastian},
  booktitle = 	 {Proceedings of the 6th Annual Learning for Dynamics \& Control Conference},
  pages = 	 {349--360},
  year = 	 {2024},
  editor = 	 {Abate, Alessandro and Cannon, Mark and Margellos, Kostas and Papachristodoulou, Antonis},
  volume = 	 {242},
  series = 	 {Proceedings of Machine Learning Research},
  month = 	 {15--17 Jul},
  publisher =    {PMLR},
  url = 	 {https://proceedings.mlr.press/v242/hose24a.html}
}

@article{kim2025bridging,
   title={Bridging the Sim-to-Real Gap: A Control Framework for Imitation Learning of Model Predictive Control},
   volume={24},
   ISSN={2005-4092},
   url={http://dx.doi.org/10.1007/s12555-026-00040-7},
   DOI={10.1007/s12555-026-00040-7},
   number={3},
   journal={International Journal of Control, Automation, and Systems},
   publisher={Springer Science and Business Media LLC},
   author={Kim, Seungtaek and Lee, Jonghyup and Han, Kyoungseok and Choi, Seibum B.},
   year={2026},
   month=Feb, pages={425–443} 
}

@ARTICLE{tagliabue2024efficient,
  author={Tagliabue, Andrea and How, Jonathan P.},
  journal={IEEE Transactions on Robotics}, 
  title={Efficient Deep Learning of Robust Policies From MPC Using Imitation and Tube-Guided Data Augmentation}, 
  year={2024},
  volume={40},
  number={},
  pages={4301-4321},
  doi={10.1109/TRO.2024.3431988}
}

@ARTICLE{ames2017control,
  author={Ames, Aaron D. and Xu, Xiangru and Grizzle, Jessy W. and Tabuada, Paulo},
  journal={IEEE Transactions on Automatic Control}, 
  title={Control Barrier Function Based Quadratic Programs for Safety Critical Systems}, 
  year={2017},
  volume={62},
  number={8},
  pages={3861-3876},
  doi={10.1109/TAC.2016.2638961}
}

@article{mohanty2025embedded,
    author = {Mohanty, Nirlipta Ranjan and Ubare, Promod and Patne, Vaishali and Ingole, Deepak and Sonawane, Dayaram},
    title = {Embedded DNN-based nonlinear MPC with guaranteed constraint satisfaction},
    year = {2025},
    issue_date = {Jun 2025},
    publisher = {Springer-Verlag},
    address = {Berlin, Heidelberg},
    volume = {37},
    number = {18},
    issn = {0941-0643},
    url = {https://doi.org/10.1007/s00521-025-11063-2},
    doi = {10.1007/s00521-025-11063-2},
    journal = {Neural Computing and Applications},
    month = mar,
    pages = {12305–12328},
    numpages = {24}
}

@INPROCEEDINGS{adhau2021constrained,
  author={Adhau, Saket and Naik, Vihangkumar V. and Skogestad, Sigurd},
  booktitle={2021 60th IEEE Conference on Decision and Control (CDC)}, 
  title={Constrained Neural Networks for Approximate Nonlinear Model Predictive Control}, 
  year={2021},
  volume={},
  number={},
  pages={295-300},
  doi={10.1109/CDC45484.2021.9683320}
}

@INPROCEEDINGS{vanroye2023fatrop,
  author={Vanroye, Lander and Sathya, Ajay and De Schutter, Joris and Decré, Wilm},
  booktitle={2023 IEEE/RSJ International Conference on Intelligent Robots and Systems (IROS)}, 
  title={FATROP: A Fast Constrained Optimal Control Problem Solver for Robot Trajectory Optimization and Control}, 
  year={2023},
  volume={},
  number={},
  pages={10036-10043},
  doi={10.1109/IROS55552.2023.10342336}
}

@ARTICLE{Lembono2020memory,
  author={Lembono, Teguh Santoso and Paolillo, Antonio and Pignat, Emmanuel and Calinon, Sylvain},
  journal={IEEE Robotics and Automation Letters}, 
  title={Memory of Motion for Warm-Starting Trajectory Optimization}, 
  year={2020},
  volume={5},
  number={2},
  pages={2594-2601},
  doi={10.1109/LRA.2020.2972893}
}

@article{Florez2023impact,
   title={IMPACT: A Toolchain for Nonlinear Model Predictive Control Specification, Prototyping, and Deployment},
   volume={56},
   ISSN={2405-8963},
   url={http://dx.doi.org/10.1016/j.ifacol.2023.10.1451},
   DOI={10.1016/j.ifacol.2023.10.1451},
   number={2},
   journal={IFAC-PapersOnLine},
   publisher={Elsevier BV},
   author={Florez, Alvaro and Astudillo, Alejandro and Decré, Wilm and Swevers, Jan and Gillis, Joris},
   year={2023},
   pages={3164–3169} 
}

@INPROCEEDINGS{Florez2025practical,
  author={Florez, Alvaro and Decré, Wilm and Swevers, Jan},
  booktitle={2025 European Control Conference (ECC)}, 
  title={Practical Approaches for Time-Optimal MPC: Dual Stage and Artificial Reference Methods}, 
  year={2025},
  volume={},
  number={},
  pages={358-363},
  doi={10.23919/ECC65951.2025.11187300}
}

@INPROCEEDINGS{Verschueren2017Time,
  author={Verschueren, Robin and Ferreau, Hans Joachim and Zanarini, Alessandro and Mercangöz, Mehmet and Diehl, Moritz},
  booktitle={2017 IEEE 56th Annual Conference on Decision and Control (CDC)}, 
  title={A stabilizing nonlinear model predictive control scheme for time-optimal point-to-point motions}, 
  year={2017},
  volume={},
  number={},
  pages={2525-2530},
  doi={10.1109/CDC.2017.8264024}
}

@inproceedings{bain2000aframework,
    author = {Bain, Michael and Sammut, Claude},
    title = {A Framework for Behavioural Cloning},
    year = {1999},
    isbn = {0198538677},
    publisher = {Oxford University},
    address = {GBR},
    booktitle = {Machine Intelligence 15, Intelligent Agents [St. Catherine's College, Oxford, July 1995]},
    pages = {103–129},
    numpages = {27}
}

@InProceedings{ross2010efficient,
  title = 	 {A Reduction of Imitation Learning and Structured Prediction to No-Regret Online Learning},
  author = 	 {Ross, Stephane and Gordon, Geoffrey and Bagnell, Drew},
  booktitle = 	 {Proceedings of the Fourteenth International Conference on Artificial Intelligence and Statistics},
  pages = 	 {627--635},
  year = 	 {2011},
  editor = 	 {Gordon, Geoffrey and Dunson, David and Dudík, Miroslav},
  volume = 	 {15},
  series = 	 {Proceedings of Machine Learning Research},
  address = 	 {Fort Lauderdale, FL, USA},
  month = 	 {11--13 Apr},
  publisher =    {PMLR},
  url = 	 {https://proceedings.mlr.press/v15/ross11a.html},
}

@InProceedings{gal2016dropout,
  title = 	 {Dropout as a Bayesian Approximation: Representing Model Uncertainty in Deep Learning},
  author = 	 {Gal, Yarin and Ghahramani, Zoubin},
  booktitle = 	 {Proceedings of The 33rd International Conference on Machine Learning},
  pages = 	 {1050--1059},
  year = 	 {2016},
  editor = 	 {Balcan, Maria Florina and Weinberger, Kilian Q.},
  volume = 	 {48},
  series = 	 {Proceedings of Machine Learning Research},
  address = 	 {New York, New York, USA},
  month = 	 {20--22 Jun},
  publisher =    {PMLR},
  url = 	 {https://proceedings.mlr.press/v48/gal16.html}
}

@inproceedings{lakshminarayanan2017simple,
    author = {Lakshminarayanan, Balaji and Pritzel, Alexander and Blundell, Charles},
    title = {Simple and scalable predictive uncertainty estimation using deep ensembles},
    year = {2017},
    isbn = {9781510860964},
    publisher = {Curran Associates Inc.},
    address = {Red Hook, NY, USA},
    booktitle = {Proceedings of the 31st International Conference on Neural Information Processing Systems},
    pages = {6405–6416},
    numpages = {12},
    location = {Long Beach, California, USA},
    series = {NIPS'17}
}

@InProceedings{laskey2017dart,
  title = 	 {DART: Noise Injection for Robust Imitation Learning},
  author = 	 {Laskey, Michael and Lee, Jonathan and Fox, Roy and Dragan, Anca and Goldberg, Ken},
  booktitle = 	 {Proceedings of the 1st Annual Conference on Robot Learning},
  pages = 	 {143--156},
  year = 	 {2017},
  editor = 	 {Levine, Sergey and Vanhoucke, Vincent and Goldberg, Ken},
  volume = 	 {78},
  series = 	 {Proceedings of Machine Learning Research},
  month = 	 {13--15 Nov},
  publisher =    {PMLR},
  url = 	 {https://proceedings.mlr.press/v78/laskey17a.html}
}

@book{Slotine1991AppliedNC,
  title     = {Applied Nonlinear Control},
  author    = {Slotine, Jean-Jacques E. and Li, Weiping},
  year      = {1991},
  publisher = {Prentice-Hall},
  address   = {Englewood Cliffs, NJ},
  isbn      = {0130408905}
}

@article{li2026neural,
  author  = {Li, Tianxun and You, Keyou},
  title        = {Neural MPC for quadrotor trajectory tracking on embedded GPUs},
  journal      = {Control Theory and Technology},
  year         = {2026},
  volume       = {24},
  pages        = {271--281},
  doi          = {10.1007/s11768-025-00304-3},
  url          = {https://doi.org/10.1007/s11768-025-00304-3},
  publisher    = {Springer},
  received     = {2025-08-01},
  revised      = {2025-09-10},
  accepted     = {2025-09-25},
  published    = {2026-02-19},
}

@ARTICLE{kong2025annmpc,
  author={Kong, Chuixu and Wang, Bin and He, Liang and Jiang, Qinglong and Liu, Chang},
  journal={IEEE Robotics and Automation Letters}, 
  title={ANN-MPC: An Adaptive Neural Network-Based Nonlinear Model Predictive Control Method for Robust Trajectory Tracking of Quadrotors}, 
  year={2025},
  volume={10},
  number={10},
  pages={10966-10973},
  doi={10.1109/LRA.2025.3607272}
}

@inproceedings{Morimasa2023approximation,
  author    = {Okamoto, Morimasa and Ren, Jiayang and Mao, Qiangqiang and Cao, Yankai},
  title     = {Approximation of nonlinear model predictive control using mixture density networks},
  booktitle = {Proc. Found. Comput.-Aided Process Oper./Chem. Process Control (FOCAPO/CPC)},
  year      = {2023}
}

@ARTICLE{karg2020efficient,
  author={Karg, Benjamin and Lucia, Sergio},
  journal={IEEE Transactions on Cybernetics}, 
  title={Efficient Representation and Approximation of Model Predictive Control Laws via Deep Learning}, 
  year={2020},
  volume={50},
  number={9},
  pages={3866-3878},
  doi={10.1109/TCYB.2020.2999556}
}

@article{alberto2002explicit,
    title = {The explicit linear quadratic regulator for constrained systems},
    journal = {Automatica},
    volume = {38},
    number = {1},
    pages = {3-20},
    year = {2002},
    issn = {0005-1098},
    doi = {https://doi.org/10.1016/S0005-1098(01)00174-1},
    url = {https://www.sciencedirect.com/science/article/pii/S0005109801001741},
    author = {Alberto Bemporad and Manfred Morari and Vivek Dua and Efstratios N. Pistikopoulos}
}

@article{zieger2020towards,
    title = {Towards Safe Neural Network Supported Model Predictive Control},
    journal = {IFAC-PapersOnLine},
    volume = {53},
    number = {2},
    pages = {5246-5251},
    year = {2020},
    note = {21st IFAC World Congress},
    issn = {2405-8963},
    doi = {https://doi.org/10.1016/j.ifacol.2020.12.1201},
    url = {https://www.sciencedirect.com/science/article/pii/S2405896320315962},
    author = {T. Zieger and A. Savchenko and T. Oehlschlägel and R. Findeisen},
}

@misc{bos2026impact,
  author       = {Bos, Mathias and Decr{\'e}, Wilm and Swevers, Jan},
  title        = {{ImpaC++t}: A {C++} Toolchain for Seamless Deployment of
                  Nonlinear Model Predictive Control with {Impact}},
  howpublished = {Presented at the 45th Benelux Meeting on Systems and Control,
                  Lommel, Belgium},
  month        = mar,
  year         = {2026},
  note         = {Abstract in the Book of Abstracts},
  url          = {https://beneluxmeeting.be/2026/}
}

@article{hsu2024safety,
  title   = {The Safety Filter: A Unified View of Safety-Critical Control in Autonomous Systems},
  author  = {Hsu, Kai-Chieh and Hu, Haimin and Fisac, Jaime F.},
  journal = {Annual Review of Control, Robotics, and Autonomous Systems},
  volume  = {7},
  number  = {1},
  pages   = {47--72},
  year    = {2024},
  doi     = {10.1146/annurev-control-071723-102940}
}

@article{christiano2016transfer,
  author  = {Christiano, Paul and Shah, Zain and Mordatch, Igor and Schneider, Jonas and Blackwell, Trevor and Tobin, Joshua and Abbeel, Pieter and Zaremba, Wojciech},
  title   = {Transfer from Simulation to Real World through Learning Deep Inverse Dynamics Model},
  journal = {arXiv preprint arXiv:1610.03518},
  year    = {2016}
}

@inproceedings{giusti2026stride,
  author    = {Giusti, Andrea and Monsieur, M. and Mrak, B.},
  title     = {{STRIDE}: Investigating Efficient Sampling Strategies for Neural Approximations of Optimal Control},
  booktitle = {Proceedings of the 23rd International Conference on Informatics in Control, Automation and Robotics (ICINCO)},
  volume    = {1},
  pages     = {166--173},
  year      = {2026},
  isbn      = {978-989-758-870-9},
  issn      = {2184-2809}
}

\end{document}